\pdfoutput=1
\documentclass[12pt]{iopart}
\usepackage{cite,amssymb,amsfonts,graphicx,bm,dcolumn,mathrsfs}
\usepackage[utf8]{inputenc}

\begin{document}

\title{Nonequilibrium Phenomena in Confined Systems}

\author{Giancarlo Franzese $^{1}$, Ivan Latella $^{2}$ and J. Miguel Rubi $^{3,}$*}

\address{$^{1}$ \quad Secció de Física Estadística i Interdisciplinària - Departament de Física de la Matèria Condensada, Facultat de Física,
  \& Institute of Nanoscience and Nanotechnology (IN2UB), 
    Universitat de Barcelona, Martí i Franquès 1, Barcelona 08028, Spain; gfranzese@ub.edu}
    
\address{ $^{2}$ \quad Laboratoire Charles Fabry, UMR 8501, Institut d'Optique, CNRS, Universit\'{e} Paris-Saclay, 2 Avenue Augustin Fresnel, 91127 Palaiseau Cedex, France; ivan.latella@institutoptique.fr}

\address{$^{3}$ \quad Secció de Física Estadística i Interdisciplinària - Departament de Física de la Matèria Condensada, Facultat de Física, Universitat de Barcelona, Martí i Franquès 1, 08028 Barcelona, Spain}




\maketitle

Confined systems exhibit a large variety of nonequilibrium
phenomena. In this special issue we have collected a limited number of
papers that where presented during the XXV Sitges Conference on Statistical
Mechanics, devoted to "Nonequilibrium phenomena in confined
systems''. The conference took place in Barcelona from the 6th until
the 10th of June 2016 (http://www.ffn.ub.es/~sitges25/), organized by
G. Franzese, I. Latella, D. Reguera and J.M. Rubi,
and gathered more than 60 international scientists in the areas of physics, chemistry and biology
working on confined systems in topics like: Diffusion and entropic transport in
confined systems; Ion and polymer translocation; Phase transitions and chemical
reactions in confined media; forces induced by fluctuations in confined systems
and Casimir effect; Confined active mater; Macromolecular crowding; Energy
conversion in confinement.

In the first contribution to this special issue~\cite{paper1}, by
P. Malgaretti, I. Pagonabarraga and J.M. Rubi, the authors focus on
how local forces in heterogeneous media modify Brownian motion and leads
to deviations in the 
Gaussian probability distribution of displacements typical of thermal motion. Their
results can be used to detect local forces and to characterize
relevant properties of the host medium.

Crowding is another source of heterogeneous local forces. P.M. Blanco,
M. Via, J.L. Garcés, S. Madurga and F. Mas analyze in their contribution~\cite{paper7} 
protein diffusion in crowded media. They use dextran macromolecules as
obstacles and propose a model based on effective radii accounting for
macromolecular compression induced by crowding. They adopt a Brownian
dynamics computational model to calculate the diffusion coefficient
and the anomalous diffusion exponent. They compare 
their results with experiments, emphasizing the effects of varying
volume fraction and hydrodynamic interactions.

The search for the origin of life motivates the study performed by
D. Niether and S. Wiegand~\cite{paper3} about the 
accumulation process of formamide in hydrothermal pores.
The authors consider a heuristic approach and show that the 
combination of thermophoresis and convection in hydrothermal pores
leads to accumulation of formamide that depends on the geometry of the system and
ambient conditions.  A sufficiently high  concentration of formamide could allow the
synthesis of prebiotic molecules. 

Smaller bio-pores are considered by M. Aguilella-Arzo,
M. Queralt-Martín, M.-L. Lopez and A. Alcaraz that in their
contribution~\cite{paper8}
address fluctuation-driven transport of ions in 
nanopores. The authors study the bacterial channel OmpF under
conditions similar to those met in vivo.  They use a three-dimensional
structure-based theoretical approach to shed light on
the conditions to observe the actual transport of ions against their
concentration gradient. 

Active matter is another biological relevant subject on which focuses 
 the contribution by A. Geiseler, P. Hänggi and
F. Marchesoni~\cite{paper6}. The authors consider the taxis of
artificial swimmers,  a purely stochastic effect associated with a
non-uniform activation of the particles’ self-propulsion. They study,
with numerical and analytical techniques, how  such swimmers respond
to a spatio-temporally modulated activating medium.

A  stochastic process is also at the base of the 
Feynman-Smoluchowski ratchet introduced 
and solved by V. Holubec, A. Ryabov, M.H. Yaghoubi,
M. Varga, A. Khodaee, M.E. Foulaadvand and P. Chvosta in their
contribution~\cite{paper9}. Using a generalization of the Fick-Jacobs 
theory, they demonstrate a
connection between the ratchet effect emerging in the model and the
rotations of the probability current of particles. In addition, the direction of
the particles mean velocity is explained using a simple discrete
analogue of the model.  

A comparison between biological and inorganic confinement is provided
in the contribution by G. Camisasca, M. De Marzio, M. Rovere and
P. Gallo~\cite{paper12}.
The authors perform numerical simulations in two situations: water confined in a hydrophilic
pore that mimics an MCM-41 (inorganic) environment and water at
interface with a protein.  They address the  slow dynamics and structural changes of
supercooled water under these confinements and  compare how the 
$\alpha$-relaxation changes with respect to bulk water. 

The simulations of J. Martí, C. Calero and G. Franzese focus, instead, only on 
water under inorganic confinement~\cite{paper10}. They
analyze the structure and dynamics of water at the 
 interface of different carbon-based materials, including armchair
 carbon nanotubes and a variety of graphene sheets (flat and with
 corrugation). They show that the diffusion of water confined between
 parallel walls depends on the plate distance in a non-monotonic way
 and is related to the water structuring, crystallization, re-melting
 and evaporation for decreasing inter-plate distance. 

 Structural, thermal, and dynamical behaviors of ionic liquids
 confined in silica ionogels are analyzed by neutron scattering in the
 experiments of  
 S. Mitra, C. Cerclier, Q. Berrod, F. Ferdeghini, R. de
 Oliveira-Silva, P. Judeinstein, J. le Bideau and J.-M. Zanotti.
  In their contribution~\cite{paper11} the authors 
 discuss various dynamic parameters and the detailed nature of phase
 transitions at  time- and length-scales smaller
 than those of  earlier experiments  with different 
  techniques. In particular, their results explain
 the good ionic conductivity of ionogels. 

A kinetic theory of a confined quasi-two-dimensional gas of hard
spheres is presented in the contribution by J.J. Brey,
V. Buzón, M.I. García de Soria and P. Maynar~\cite{paper4}.
In this paper, a model
formulated for granular gases is reviewed and a Boltzmann kinetic
equation for elastic hard spheres is introduced in order to obtain a
detailed description of the dynamics of the system. 

 Another inorganic system particularly interesting
 is that  made of a few confined colloids.  
 I.A. Martínez, C. Devailly, A. Petrosyan and S. Ciliberto discuss
 energy transfer between colloids via critical
 interactions~\cite{paper5}. By considering two beads that are trapped
 by two optical tweezers whose distance is periodically modulated,
 they report a temperature-controlled synchronization of the particles
 in a binary mixture close to the critical point of the demixing
 transition. 

 Noise can induce also coherence at the level of electron spins.
 B. Spagnolo, C. Guarcello, L. Magazzù,
 A. Carollo, D. Persano Adorno and D. Valenti
 investigate 
 nonlinear relaxation phenomena in various metastable condensed matter
systems~\cite{paper2}: phase dynamics in Josephson
junctions, electron spin relaxation in an n-type GaAs bulk driven by a
fluctuating electric field and stabilization of quantum metastable
states by dissipation.
They
discuss with detail the
relevance of noise enhanced stability, stochastic resonant activation
and the noise-induced coherence of electron spin in these systems.

With the organization of the Conference we aimed to provide a unique
opportunity to exchange points of view, to promote contacts and new
collaborations among researchers working on different inter-disciplinary areas,
and to create a forum for debate that could help to answer the open
questions related to
the novel effects induced by confinement. This issue is the natural
continuation of the interchange of ideas that started during the Conference
sessions. We hope that this small selection of contributions will
trigger further investigation and discussion among their readers.



We would like to express our sincere gratitude to the authors of the papers that made possible this special issue and to the editorial team of Entropy for priceless assistance during the process.

The authors declare no conflict of interest.

\end{document}